\documentstyle[epsf]{mn}
\def\sec{$^{\prime\prime}$} 
\def\deg{$^\circ$\ }

\newcommand{\lesssim}{\mathrel{\hbox{\rlap{\hbox{\lower4pt\hbox{$\sim$}}}\hbox{$<$}}}}
\newcommand{\gtrsim}{\mathrel{\hbox{\rlap{\hbox{\lower4pt\hbox{$\sim$}}}\hbox{$>$}}}}
\begin{document}

\title{Extended gas in Seyfert galaxies: Near infrared observations of 
NGC\,2110 and Circinus}

\author[T. Storchi-Bergmann et al.]{Thaisa Storchi-Bergmann$^1$\thanks
{Visiting Astronomer at the Cerro Tololo Interamerican Observatory, 
operated by the Association of Universities for Research in Astronomy, 
Inc. under contract with the National Science Foundation.}, 
Cl\'audia Winge$^{1,2\star}$, Martin J. Ward$^{3\star}$ and \newauthor Andrew S. Wilson$^{4,2\star}$ \\
$^1$ Instituto de F\'\i sica, UFRGS,
Campus do Vale, C.P. 15051, P. Alegre, RS, BRAZIL\\
$^2$ Space Telescope Science Institute, 3700 San Martin Drive, 
Baltimore, MD\,21218, USA\\
$^3$ Dept. of Physics and Astronomy, University of Leicester,
University Road, Leicester LE1 7RH, England\\
$^4$ Astronomy Department, University of Maryland,
College Park, MD\,20742, USA}
 
\maketitle

\begin{abstract}

We present results of near--IR long-slit spectroscopy
in the J and K bands of
the Seyfert 2 galaxies NGC\,2110 and Circinus. 
Our goal is to investigate the gaseous distribution,
excitation, reddening and kinematics, looking for signatures
of the molecular torus hypothesised in unified models
to both obscure and collimate the nuclear radiation.
The two galaxies show extended emission in the IR emission lines
[Fe\,II]$\lambda$1.257$\mu$m, Pa$\,\beta$ and H$_{2} v$=1--0 S(1),
both along the major axis of the galaxy disk and perpendicular to it. 
In NGC 2110, the emission line ratio [Fe\,II]/Pa$\,\beta$
increases towards the nucleus, where its value is
$\approx 7$. Further, the nuclear [Fe II] and Pa$\,\beta$ lines are
broader (FWHM $\simeq$ 500 km s$^{-1}$) than the H$_2$ line
(FWHM $\le$ 300 km s$^{-1}$). Both these results suggest that
shocks, driven by the radio jet, are an important source of
excitation of [Fe II], while the H$_2$ excitation is dominated by
X-rays from the nucleus. Br$\,\gamma$ is only observed at the nucleus, where H$_2$/Br$\,\gamma \approx 3$. 
In the case of Circinus, both [Fe\,II]/Pa$\,\beta$ and 
H$_2$/Br$\,\gamma$ decrease from $\approx$ 2 at 4\sec\ from the nucleus
to nuclear values of $\approx\,$0.6 and $\approx\,$1, respectively, 
suggesting that the starburst dominates the nuclear excitation, 
while the AGN dominates the excitation further out (r$\,\ge $2\sec). 
For both galaxies, the gaseous kinematics are 
consistent with circular rotation in the plane of the disk.
Our rotation curves suggest that the nucleus (identified with the peak
of the IR continuum) is displaced from the kinematic centre of the galaxies.
This effect has been observed previously in NGC\,2110 based on the kinematics
of optical emission lines, but the displacement is smaller in the infrared,
suggesting the effect is related to obscuration.
The continuum J--K colours indicate a red stellar population in 
NGC\,2110 and a reddened young stellar population in Circinus, outside the
nucleus. Right at the nucleus of both galaxies, the colours are redder,
apparently a result of hot dust emission, 
perhaps from the inner edge of a circumnuclear torus. 
In NGC\,2110, the signature of the hot dust emission is particularly clear
in the K--band, being seen as an additional component superimposed 
on the continuum observed in the J--band. 

\end{abstract}


\section{Introduction}

The nuclear regions of Seyfert 2 galaxies are dusty environments,
as revealed by red nuclear colours in optical imaging and
large Balmer emission-line ratios in spectroscopic studies 
(e.g. Ward et al. 1987; Storchi-Bergmann, Wilson \& Baldwin 1992;
Mulchaey et al. 1994, Storchi-Bergmann, Kinney \& Challis 1995;
Simpson et al. 1996; Cid-Fernandes,
Storchi-Bergmann \& Schmitt 1998). Although such reddening
may be due to large scale dust in the central regions of the
host galaxies, dust on smaller scales is believed to be concentrated in   
an optically thick molecular torus 
with dimensions of tens of parsecs and surrounding the nuclear engine
(e.g. Antonucci \& Miller 1985; Antonucci 1993).
This torus both obscures the nucleus from direct view and 
collimates the ionizing radiation, giving origin to the ``ionization
cones'' observed in Seyfert 2 galaxies (e.g. 
Mulchaey, Wilson \& Tsvetanov 1996). Evidence for reradiation from hot
dust emitting in the near-IR has been found, among others, by
McAlary \& Rieke (1988); Sanders et al. (1989); 
and Alonso-Herrero, Ward \& Kotilainen (1996).

In order to penetrate the dust layers in Seyfert 2 galaxies,
it is necessary to observe in the infrared region of the spectrum.
At $\lambda\approx 2\mu$m, $A_K \approx A_V/10$, so near-IR observations can
reach deeper into the nuclear region than optical observations.
In addition, if there is warm molecular hydrogen near the nucleus,
emission from the vibration-rotational transitions of molecular
hydrogen (such as H$_2\,v$=1--0 S(1) at rest wavelength $2.122\,\mu$m)
should be strong. If an H$_2$--emitting disc--like structure could be detected
elongated perpendicular to the ionization cone or radio axis, 
this would provide very strong support for the unified model.
Narrow-band imaging studies by Blietz {\it et al.\ }(1994) 
have indeed shown that in NGC~1068 the H$_2$ line emission is
spatially extended almost perpendicular to the cone.
Similar results have been found for NGC~4945 (Moorwood et al. 1996).

With these issues in mind, we have obtained long-slit spectra in the
J and K--bands of a number of Seyferts with anisotropic
high-excitation optical emission, which, in the unified model,
is a result of ionization and excitation by
nuclear radiation collimated by the torus. We present in this work
the results for NGC\,2110 and Circinus, the two galaxies for which 
we were able to detect IR emission lines
at the largest distances from the nuclei.

NGC\,2110 is an early-type Seyfert 2 galaxy discovered 
through its X-ray emission (Bradt et al. 1978).
Narrow-band images show high excitation gas extending up to 
10$^{\prime\prime}$ from the nucleus 
(Wilson, Baldwin \& Ulvestad 1985; Mulchaey et al. 1994),
in a similar direction to a jet--like radio source at
position angle (p.a.) $10^\circ$ (Ulvestad \& Wilson 1983).
Wilson \& Baldwin (1985) and Wilson, Baldwin \& Ulvestad (1985),
using optical long-slit spectroscopy, have
found extended ionized gas in normal rotation about 
the photometric minor axis, but
with the kinematic centre displaced $\approx 1.7^{\prime\prime}$
south of the optical continuum nucleus. Adopting a distance to 
NGC\,2110 of 31.2\,Mpc (using the systemic velocity from
Wilson \& Baldwin 1985, for H$_0$ = 75 km~s$^{-1}$), we obtain a scale of
151\,pc\,arcsec$^{-1}$.

The Circinus galaxy is a nearby ($\sim$ 4 Mpc) 
spiral (Freeman et al. 1977) with 
a nuclear spectrum characteristic of
both starburst and Seyfert activity (Moorwood \& Oliva 1988, 1990).
It contains the closest known ionization cone and a circumnuclear starburst ring
(Marconi et al. 1994). It is a strong H$_2$O megamaser emitter, 
and shows radio lobes approximately orthogonal to the galactic plane 
(Elmoutie et al. 1995). Maiolino et al. (1998) present H and K images
of the inner 13\sec$\times$13\sec,
narrow-band images of the central 5\sec$\times$5\sec and a study of the stellar kinematics using near-IR absorption bands. Davies et al. (1998) present
narrow-band images in [Fe\,II]$\lambda$1.64$\mu$m and
H$_2\,v$=1--0 S(1), as well as radio continuum maps at 3 and 6 cm.
Our adopted distance of 4\,Mpc (Freeman et al. 1977) corresponds
to a scale of 19\,pc\,arcsec$^{-1}$.

\section{Observations}

Long-slit spectra of NGC\,2110 and Circinus in the J and K bands
were obtained using the Infrared Spectrograph (IRS) on the 4m telescope 
of the Cerro Tololo Interamerican Observatory in the nights of Nov. 1 and 2,
1995 and March 2, 1996. The scale of the 256$\times$256 InSb 
detector was 0.363\sec per pixel, and the useful
slit length was 15$^{\prime\prime}$.
Two gratings were used: one with 75~lines per mm,
and resolution $R\approx 700$ ($\sim$4 pixels), hereafter LR grating, 
and the other with 210~lines per mm, 
and $R\approx 2000$, hereafter HR grating. The corresponding
velocity resolutions are $\approx$ 400 km s$^{-1}$ and 150 km s$^{-1}$,
respectively. The slit width was either 1.1$^{\prime\prime}$ or 1.7$^{\prime\prime}$, depending on the seeing. 

NGC\,2110 was observed along two position angles: p.a.\,=\,$170^\circ$, which 
is close to both the major axis of the 
inner isophotes in the narrow--band [OIII]
image of Mulchaey et al. (1994) and the
major axis of the galaxy disk, and along the perpendicular direction,
at p.a.\,=\,80$^\circ$. 
Circinus was observed along the approximate axis of the
radio lobes in p.a.\,=\,$-$66$^\circ$, and
along the approximate galaxy major axis in p.a.\,=\,24$^\circ$.
A log of the observations is presented in Table 1.

\begin{table*}
\begin{minipage}{150mm}
\caption{Log of Observations}
\begin{tabular}{lcccccc}
\hline\hline
Object    &Date  &P.A.(\deg) &Band ($\mu$m)  &Grating &Exp.Time (sec)  &Slit width(\sec)\\ 
\hline
NGC\,2110 &1 November 1995 &170 &J(1.24-1.35) &75 l/mm &2400 &1.1\\
          &2 November 1995 &80 &K(2.07-2.24) &75 l/mm &2160 &1.1\\
          &2 November 1995 &170 &K(2.07-2.24) &75 l/mm &1800 &1.1\\
          &2 March 1996 &170 &J(1.26-1.30) &210 l/mm &1800 &1.7\\
          &&&&&&\\
Circinus  &2 March 1996 &24 &J(1.24-1.29) &210 l/mm &1800 &1.7\\
          &2 March 1996 &$-$66 &J(1.24-1.29) &210 l/mm &800 &1.7\\
          &2 March 1996 &$-$66 &K(2.12-2.18) &210 l/mm &800 &1.7\\
          &2 March 1996 &24  &K(2.12-2.18) &210 l/mm &800 &1.7\\         
\hline
\end{tabular}
\end{minipage}
\end{table*}

Wavelength calibrations in the J and K bands were obtained using
an He\,Ar lamp and OH sky lines, respectively. 
The reduction was performed using IRAF through scripts
kindly made available by Richard Elston (available at the CTIO ftp archive).
One--dimensional spectra were extracted binning 
together 2 pixels (0.73\sec), except
for the outermost locations where 
4 pixels (1.46\sec) were co--added to improve the signal to noise ratio\,(S/N). 
The atmospheric absorption features were removed using spectra of nearby stars bracketing the galaxy observations; each
extracted spectrum was divided by the normalized spectrum of
a star (or the average between the spectra of two stars, one observed before
and the other after the galaxy). 
The spectra were then flux calibrated using 
observations of standard stars from Elias et al. (1982).

\section{Results}

\subsection{NGC\,2110}

\subsubsection{Emission-line Profiles}

Sequences of J and K--band spectra obtained with the LR
grating are presented in Figs. 1 (p.a.\,=\,170$^\circ$) and 2 (p.a.\,=\,80$^\circ$). The higher dispersion spectra
obtained with the HR grating are shown in Fig. 3.
The strongest emission line in the J--band is 
[Fe\,II]$\lambda 1.257 \mu$m, followed
by [Fe\,II]$\lambda$1.321 and Pa$\,\beta$. The profiles of
[Fe\,II]$\lambda$ 1.257 and Pa$\,\beta$ are better
defined in the higher dispersion spectrum of Fig. 3.
We find that Pa$\,\beta$ is blended with 
[Fe\,II]$\lambda 1.279\mu$m, which has a predicted flux of 7\% 
that of [Fe\,II]$\lambda 1.257\mu$m. Although the profile of
[Fe\,II]$\lambda 1.279\mu$m is not well defined (due to its low S/N), 
its full width at half maximum (FWHM)
is similar to that of [Fe\,II]$\lambda 1.257\mu$m, 
and the observed wavelength and 
flux agree with the expected values.

[Fe\,II]$\lambda 1.257\mu$m emission is
observed up to $\approx 5$\sec\ to the N and $\approx 3$\sec\ 
to the S, being stronger to N, where both the 
radio (Ulvestad \& Wilson 1983) and
optical high-excitation line ([OIII]) emission are stronger 
(Wilson, Baldwin \& Ulvestad 1985; Mulchaey et al. 1994).

\begin{figure}
\vspace{10.0cm}
\includegraphics{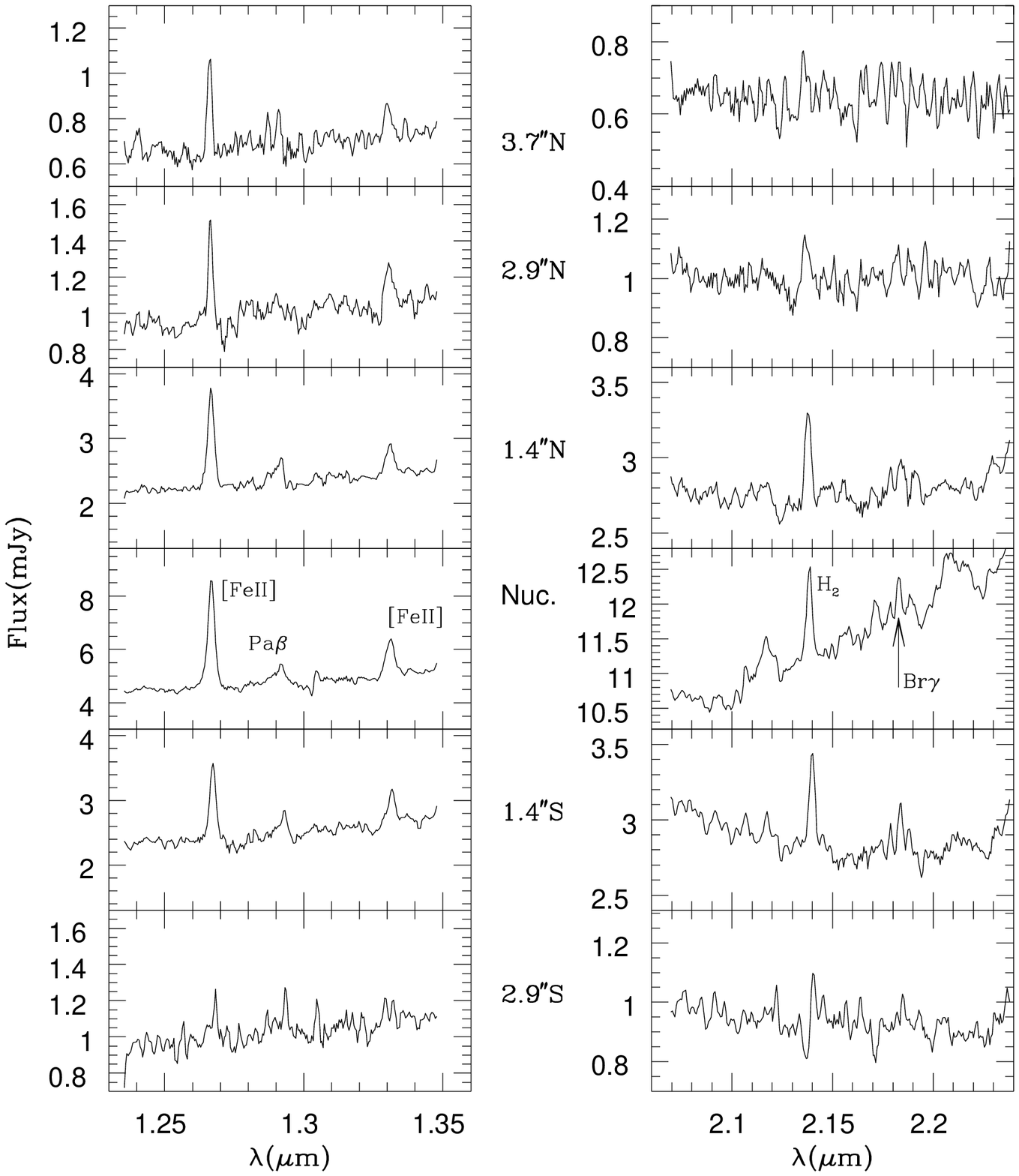}
\caption{J and K--band spectra of NGC\,2110 at $R\approx 700$
along the radio axis at p.a.\,=\,170$^\circ$, 
after binning 2 pixels (0.73\sec) together. 
The distance from the nucleus is indicated.}
\label{fig1}
\end{figure}

\begin{figure}
\vspace{9.0cm}
\includegraphics{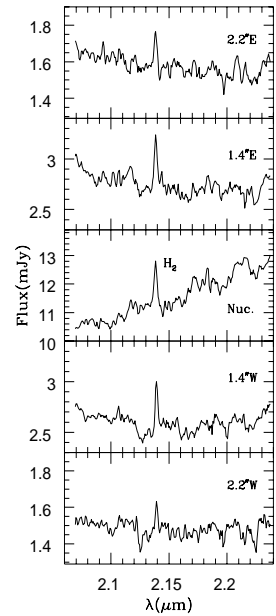}
\caption{K--band spectra of NGC\,2110 at $R\approx 700$ 
along p.a.\,=\,80$^\circ$ (perpendicular to the radio axis), 
after binning 2 pixels (0.73\sec) together. 
The distance from the nucleus is indicated.}
\label{fig2}
\end{figure}

\begin{figure}
\vspace{10.0cm}
\includegraphics{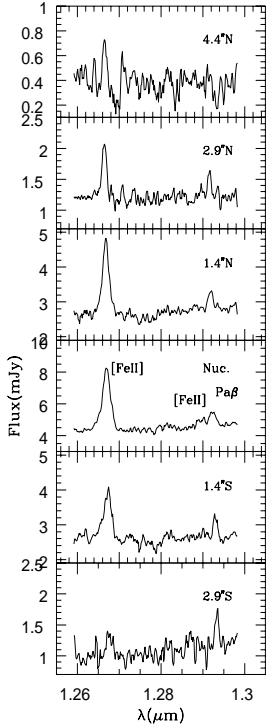}
\caption{J--band spectra of NGC\,2110 at $R\approx 2000$ 
along the radio axis at p.a.\,=\,170$^\circ$, 
after binning 2 pixels (0.73\sec) together. 
The distance from the nucleus is indicated.}
\label{fig3}
\end{figure}

In the K--band, H$_2\,v$=1--0 S(1)
is clearly detected to approximately 4\sec\ 
N and S and 3\sec\ E and W of the nucleus (Figs. 1 and 2). 
Br$\,\gamma$ is detected at the nucleus and at 1.4\sec\ S, but
the galaxy continuum is very strong and has a lot of structure,
resulting in a large uncertainty for the Br$\,\gamma$ flux. 

Our nuclear emission-line 
spectra in the J and K bands are similar to
those of Veilleux, Goodrich \& Hill (1997), which were obtained with 
a much larger aperture (3\sec $\times$ 3\sec). The difference in 
aperture size may be responsible for the 
difference in the K--band continuum, which is much redder in our data. 
An unresolved, very red nucleus is also found in
the HST optical continuum colour map of Mulchaey et al. (1994).
Veilleux et al. (1997) report a ratio betwen the narrow  
Br$\,\gamma$ and H$_2$ fluxes of
$\approx 0.25$, as compared with $\approx 0.35$ in
our case. They also 
say they may have detected broad components to Pa$\,\beta$ and Br$\,\gamma$; 
the structure in the continuum around Br$\,\gamma$ 
precludes any attempt to measure such a component in our spectra.
 
In Fig. 4 we show a comparison of the line profiles. 
The top panels show the scaled profiles from the spectra obtained with 
the HR grating. From the top left panel, it can be seen that the 
nuclear Pa$\,\beta$ profile is very similar to that of the 
[Fe\,II]$\lambda 1.257\mu$m
line, although the blue wing of Pa$\,\beta$
is contaminated by  
[Fe\,II]$\lambda 1.279\mu$m, as described above.
We do not find evidence in our data for the broad component suspected
by Veilleux et al. (1997) in Pa$\,\beta$. As our data seems to
have similar S/N ratio to theirs, one possibility is that 
contamination of the blue wing of Pa$\,\beta$ by the adjacent
[Fe\,II] line may have
been interpreted as a broad component.  

\begin{figure}
\vspace{9.0cm}
\includegraphics{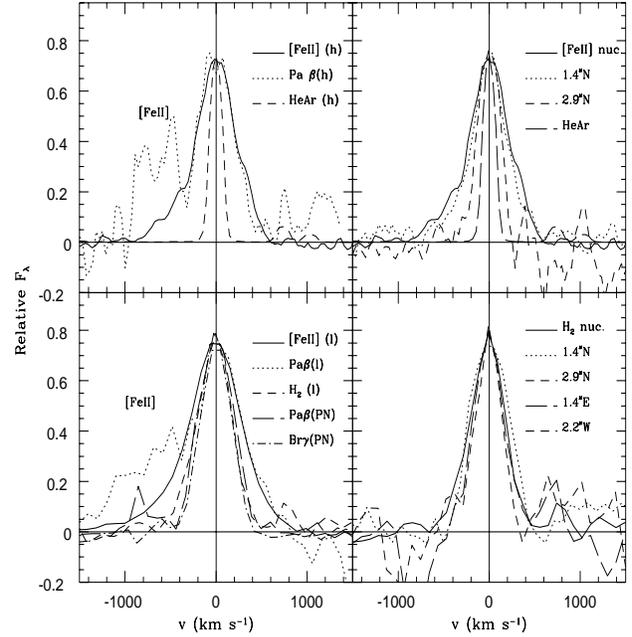}
\caption{Comparison of the emission-line profiles of NGC\,2110. The upper panels
correspond to the HR grating and the lower panels to the LR grating.
{\it Upper left:} The nuclear [Fe\,II]$\lambda$1.257$\mu$m line 
profile (continuous line) is 
compared with that of Pa$\,\beta$ (dotted line), 
as well as with the instrumental profile (dashed line). {\it Upper right:}
A comparison of the  [Fe\,II]$\lambda$1.257$\mu$m line profiles at
different distances from the nucleus along the radio axis. 
{\it Bottom left:} Comparison of the nuclear emission--line 
profiles with those of the planetary nebula NGC\,7009,
which are representative of the instrumental resolution. {\it Bottom right:}
comparison of the H$_2$ profiles as a function of distance from the nucleus.}
\label{fig4}
\end{figure}

Also in Fig. 4, the [Fe\,II] and Pa$\,\beta$ profiles
are compared with that of a He\,Ar (calibration lamp) line, which has
FWHM\,=\,150$\pm$5~km~s$^{-1}$. Although the signal-to-noise ratio
is much lower for Pa$\beta$, it can be seen that both
lines are much broader than the instrumental profile. 
The FWHM (corrected for the
instrumental FWHM above) are 520$\pm$20~km~s$^{-1}$ for [Fe\,II]
and 430$\pm$70~km~s$^{-1}$ for Pa$\beta$. The blue
and red wings of the [Fe\,II] line reach $\approx -900$~km~s$^{-1}$,
and $\approx 600$~km~s$^{-1}$ at zero intensity, respectively.
The [Fe\,II] profile
gets narrower with increasing distance
from the nucleus (top right panel of Fig. 4), 
reaching a corrected FWHM\,=\,240$\pm$40~km~s$^{-1}$ at 2.9\sec\ N of the nucleus. 

The bottom panels of Fig. 4 show the profiles from the spectra
obtained with the LR grating, where it can be seen that the
nuclear [Fe\,II] profile is still resolved. 
The similarity of the Pa$\,\beta$ and [Fe\,II] profiles 
is confirmed by these data. The corrected FWHM's of [Fe\,II] and 
Pa$\beta$ are, respectively,
500$\pm$40~km~s$^{-1}$ and 550$\pm$120~km~s$^{-1}$,
consistent with the values from the
higher resolution spectrum. On the other hand, 
the nuclear H$_2$ profile is barely resolved,
being very similar to the profiles of the Pa$\,\beta$ and Br$\,\gamma$ 
lines of the planetary nebula NGC\,7009, observed as a reference,
and which have observed (uncorrected) FWHM $\approx$ 400~km~s$^{-1}$. 
Correcting the FWHM of the nuclear H$_2$ line
for the instrumental profile (adopted as that of the Br$\,\gamma$ line 
of NGC\,7009), the resulting FWHM is 230$\pm$70~km~s$^{-1}$.
Finally, in the bottom right panel of Fig. 4 we present the
profiles of the H$_2$ line as a function of distance from the nucleus,
which show a marginal decrease of FWHM with increasing distance
from the nucleus. 

We can thus conclude that, at the nucleus of NGC 2110, the [Fe\,II] and
H$_2$ lines originate in different material, the former line coming from more
kinematically disturbed gas. The larger widths of the [Fe\,II] line could 
result from acceleration of interstellar gas by the radio jet, which might also
be responsible for ionizing the species through the agency of shock waves
(see section 3.1.3).

\subsubsection{Velocity field}

Fig. 5 shows the heliocentric 
velocities derived from the peak wavelengths of the
[Fe\,II]$\lambda 1.257\mu$m, Pa$\,\beta$ and H$_2$ emission lines,
from both the LR and HR data. The lower panel shows the 
velocities along the radio axis at
p.a.\,=\,170$^\circ$, for all the emission lines,
and the upper panel shows the velocities along p.a.\,=\,80$^\circ$,
available only for H$_2$. The good agreement between the velocities
derived for [Fe\,II] using two different gratings 
gives us confidence in the reliability of the data.
For clarity, error bars have not been plotted in the lower panel
(the sizes are similar to those in the upper panel).
The adopted position of the nucleus corresponds to the peak
of the continuum light in the J and K bands. 

There is a small difference between the kinematics of the
gas emitting [Fe\,II] and that emitting Pa$\,\beta$ and H$_2$:
the latter two emission lines show a rotation curve well
represented by a circular rotation model (see below), while the [Fe\,II] 
rotation curve shows a shallower gradient near the nucleus. 
Although small, this difference seems to be significant (as
compared with the error bars in Fig. 5) and 
may be related to interaction of the [Fe II] emitting gas with
the radio jet, already suggested by the broadening of its
emission line profile.

\begin{figure}
\vspace{10.0cm}
\includegraphics{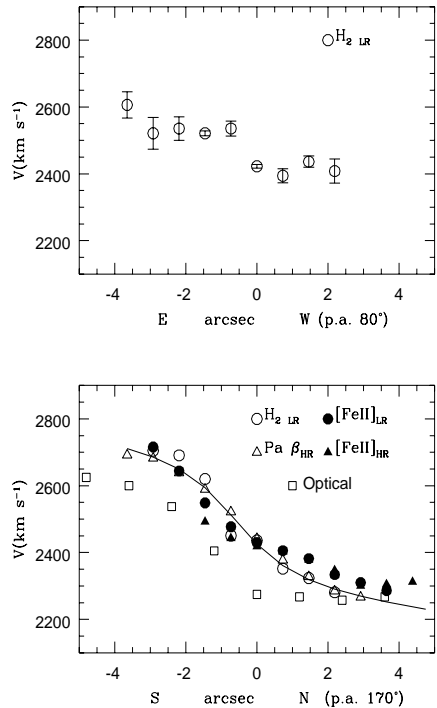}
\caption{Heliocentric rotation curve of NGC\,2110, 
obtained from the peak wavelengths
of the emission lines. The subscripts on the labels indicate the
grating used. The continuous line is the fit of a model with circular 
rotational motions to the Pa$\,\beta$ and and H$_2$ data along p.a.\,=\,170$^\circ$ (radio axis). For clarity, error bars
have been shown only in the upper panel, but are mainly dependent on
the distance to the nucleus and approximately apply also
to the lower panel.
The average peak velocities from [OIII]$\lambda$5007 and H$\beta$
(Wilson \& Baldwin 1985) are represented by open squares.}
\label{fig5}
\end{figure}

It is interesting to compare the near--IR and
optical emission-line kinematics.
Wilson \& Baldwin (1985) find that the kinematic 
centre of NGC\,2110, as derived from the optical emission-line
rotation curve, is displaced 1.7\sec\ S of the optical nucleus (identified with
the peak of continuum light) along p.a.\,=\,161$^\circ$. In order to 
search for any similar shifts
in our data, we have fitted a model
with circular rotational motions (Bertola
et al. 1991) to the peak velocities of Pa$\,\beta$ and H$_2$ 
(omitting the the [Fe\,II] line as its motions may 
be affected by interactions with the radio jet) 
along p.a.\,=\,170\deg, allowing for a shift $x_0$ 
in the position of the kinematic centre with respect to the nucleus.
This fit shows that the data are consistent 
with circular motions in the plane of the galaxy disk,
with the kinematic centre shifted by $x_0$=0.6\sec\ S with respect to the
peak of the continuum IR light. This shift is in the same direction
as that from the optical observations, but is smaller.
In order to investigate this diference, we have  
also plotted in Fig. 5 the average of  
the [OIII] and H$\beta$ peak velocities of Wilson and Baldwin (1985),
assuming the optical and infrared nuclei coincide.
The kinematic centre of this optical rotation curve is seen to be displaced
$\approx$ 1.2\sec\ S relative to the kinematic centre of the IR curve.

The above results suggest that the apparent displacement between
the kinematic and photometric nucleus is, at least in part, an effect of
obscuration.  The difference between the optical and 
IR rotation curves may result from either a shift between the IR and 
optical continuum peaks and/or the fact that the IR emission lines sample the
NLR kinematics closer to the nucleus 
(see discussion by Wilson \& Baldwin 1985
and Wilson, Baldwin \& Ulvestad 1985).

\subsubsection{Emission line fluxes and ratios}

Fig. 6 shows the line fluxes as a function of distance
from the nucleus. In order to be able to compare the flux distributions
along the the N--S and E--W directions, 
we have de-projected the angular distances assuming 
the gas lies in the plane of the stellar disk and adopting an inclination $i=53$\deg and a major axis p.a.=160\deg
(Wilson \& Baldwin 1985). The [Fe\,II]$\lambda$1.257$\mu$m 
line fluxes obtained from the LR spectra are
$\approx$ 30\% larger than that from the HR spectra, but when corrected
for the different slit widths (1.1\sec\ for the LR and 1.7\sec\ for the HR spectra), this difference increases to a factor of 2. At least part of
this difference can be attributed to a somewhat poorer seeing in
the second observing run. 

It can be seen in the left panels of Fig. 6 that the
[Fe\,II] lines are more extended to the N (5\sec) than to the 
S (3\sec), with a small ``bump" to the N, following the radio emission.
(Due to observing constraints, it was
not possbile to obtain a J spectrum perpendicular to the radio axis).
The Pa$\,\beta$ flux distribution (upper right panel of Fig. 6)
shows similar extent to N and S,
being flatter to the S. 

The H$_2$ flux distribution is presented in lower right panel
of Fig. 6. After de-projection, 
there is no significant asymmetry in either the 
N--S or E--W H$_2$ flux distributions within 3\sec\ of the
nucleus.  The H2 line is detected further from the
nucleus along the E--W direction than along the N--S direction. Although
a compact molecular torus may contribute to the H$_2$ emission, the line 
extends to $\approx 1$ kpc from the nucleus. 
Moreover, the H$_2$ kinematics are consistent
with circular motion in the disk of the galaxy (Section 3.1.2).

The luminosity in the H$_2$ line can be used to estimate the mass
of hot molecular hydrogen in NGC\,2110. We have used the method
described by Veilleux et al. (1997), which is based on 
the calculations of Scoville et al. (1982).
Adopting cylindrical symmetry for the H$_2$ flux distribution, 
we obtain an integrated H$_2\,v$=1--0 S(1)
luminosity in the inner 10\sec\ (Fig. 6)
of $L_{H_2}=6.7\times 10^{39}$ ergs s$^{-1}$.
If the hot H$_2$ molecules are thermalized at T=2000 K,
and assuming that the power in all H$_2$ lines is
10 times that in the S(1) line, this luminosity translates
into a hot H$_2$ mass of 2,300 M$_\odot$. 
If we correct the H$_2$ luminosity for extinction (adopting E(B-V)=1.5 --
see below) the luminosity and resulting mass would be $\approx$ 60\%
larger.

\begin{figure}
\vspace{10.0cm}
\includegraphics{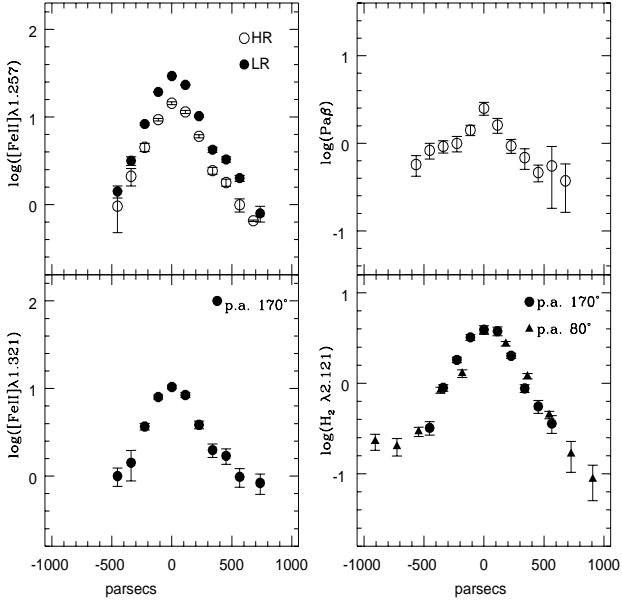}
\caption{Logarithm of the emission-line fluxes 
(in units of $10^{-15}$~ergs~cm$^{-2}$~s$^{-1}$ (arcsec)$^{-2}$,
as a function of distance from the nucleus of NGC\,2110.
Circles correspond to the direction S--N (radio axis), with S negative
and N positive. Triangles correspond to the direction E--W,
with E negative and W positive. The x axes refer to 
distances in the plane of the galaxy disk (scale: 151 pc arcsec$^{-1}$),
assuming $i=53$\deg and major axis p.a.=160\deg.
Open symbols correspond to the HR
grating, and filled symbols to the LR grating.} 
\label{fig6}
\end{figure}

The emission--line ratio [FeII]$\lambda$1.257/Pa$\,\beta$ 
as a function of distance from the nucleus
along the radio axis is presented in Fig. 7.  
It reaches very high values at the nucleus: $\approx 7$, 
much higher than the value of
$\approx 1$ found in NGC\,1068, for example (Ward et al. 1987).
It decreases more slowly to the N (along the radio emission) 
than to the S, and only in the outermost regions of NGC\,2110 does this ratio 
approach unity. 

\begin{figure}
\vspace{6.0cm}
\includegraphics{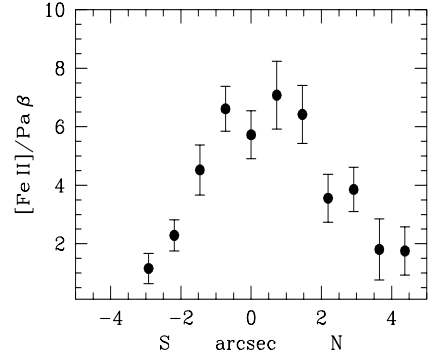}
\caption{Emission--line ratio  [FeII]$\lambda$1.257/Pa$\,\beta$
as a function of distance from the nucleus of NGC\,2110 along the radio axis at
p.a.\,=\,170\deg. Negative positions are to the S and positive to N.}
\label{fig7}
\end{figure}

The origin of the [Fe\,II] emission in Seyfert galaxies has
been extensively discussed in previous works (e.g. Forbes \& Ward 1993;
Simpson et al. 1996; Veilleux et al. 1997). 
Possibilities include ionization by X-rays from the
central source, ionization by shocks produced by interaction of
a radio jet with the surrounding medium, and contributions from starbursts.
However, detailed calculations 
by Colina (1993) show that, for starbursts, [Fe\,II]/Pa$\,\beta$~$\le 0.4$,
suggesting a starburst contribution is not important for
the nuclear region of NGC\,2110. 
Forbes \& Ward (1993) favour ionization by shocks in both starbursts and
Seyferts because of the correlation
between [Fe\,II] and radio fluxes, 
which was confirmed by the additional data of Veilleux et al. (1997).
For NGC\,2110 the broadening of the
[Fe II] profile (discussed above) at the locations where
the [FeII]$\lambda$1.257/Pa$\,\beta$ ratio is higher,
the presence of a linear radio source, and the fact that the
[Fe II] emission is more extended along the radio axis,
give further support to this scenario. The very high nuclear [FeII]$\lambda$1.257/Pa$\,\beta$ ratio can thus be understood as
due to the fact that while both Pa$\beta$ and [Fe\,II] likely contain
contributions from photoionization by the central source, the [Fe\,II]
emission is further enhanced by the shock ionization of Fe by
the radio jet.

For the nucleus, we obtain the ratio 
H$_2\,v$=1--0 S(1)/Br~$\gamma \approx 3.0\pm$ 2.0;
it was not possible to measure Br$\,\gamma$ outside the nucleus.
The ratio H$_2\,v$=1--0 S(1)/Br$\,\gamma$ allows
us to discriminate between various excitation mechanisms for the H$_2$
emission. In star-forming regions, where the main heating agent is
UV radiation, H$_2$/Br$\gamma\,<\,1.0$, while for 
Seyferts this ratio is larger due to additional H$_2$ emission 
excited by shocks or by X--rays from the active nucleus (Moorwood
\& Oliva 1994). For NGC\,2110, although some contribution from 
shocks cannot be discarded, the smaller width of the H$_2$ 
emission line when compared to that of the [Fe II]$\lambda$1.257 
suggests that the excitation of H$_2$ is dominated by
X--rays from the active nucleus.
Veilleux et al. (1997) have shown that NGC\,2110 has
enough X-ray emission to produce the necessary excitation.

\subsubsection{Reddening and Dust Emission}

The ratio between the Br$\,\gamma$ and Pa$\,\beta$ nuclear fluxes can
be used to estimate the gaseous reddening at the nucleus under 
the assumption of Case B recombination
(for T = 10$^4$ K, Br$\,\gamma$/Pa$\,\beta$~$=0.170$; Osterbrock 1989) and
the reddening law from Whitford (1958) and Rieke \& Lebofsky (1985),
through the expression:

$$  E(B-V) = 5.21 \times\log \left(\frac{\frac{F_{Br\,\gamma}}{F_{Pa\,\beta}}}
{0.170}\right) mag $$ 

\noindent
The resulting nuclear reddening, after subtraction of the
foreground Milky Way reddening E(B--V)$_G \approx$ 0.36 mag
(Burstein \& Heiles 1982) is E(B-V)$=1.1\pm 0.7$ mag.
This reddening is comparable with a previous optical determination
(E(B-V) = 0.73 mag -- Shuder 1980). The error is large
due to the uncertainties in the fluxes of 
Br$\,\gamma$ and Pa$\,\beta$ (F$_{Br\,\gamma}=1.0(\pm 0.4)\times
10^{-15}$~ergs~cm$^{-2}$s$^{-1}$; F$_{Pa \beta}=3.0(\pm 0.4)\times
10^{-15}$~ergs~cm$^{-2}$s$^{-1}$; 
both integrated over an area of 1.1\sec $\times$ 0.73 \sec).
 
The nuclear reddening can also be estimated from the slope 
of the continuum. We have used the spectra of Fig. 1 to obtain
the J--K colours: after correcting for E(B--V)$_G$, we obtain
J--K $\approx$ 1.9 mag for the nucleus and $\approx $0.9 mag for the
extranuclear spectra. The nuclear value is in good agreement with the
colours J--H$ =0.93$ and H--K$ =1.0$ obtained by Alonso-Herrero et al. (1998)
within an aperture of 1.5\sec\  diameter. 
If the intrinsic nuclear colour were the same
as that of the extranuclear regions and the observed colour
were a result of reddening of the starlight by dust, 
the nuclear obscuration
is A$_V \approx$ 6 mag, larger than  
the value obtained from the emission lines.

On the other hand,  J--K~$\approx$~2 is also consistent with a mixture
of a late-type stellar population and emission by hot dust
(Simpson et al. 1996). In order to investigate whether
obscuration or dust emission is responsible for the red nuclear colour
of NGC 2110,  we have compared the slopes of the nuclear 
continuum in our J and K band spectra with those
of the extranuclear continua reddened by various values.
In the J--band the slope of the nuclear continuum is very similar to
that of the extranuclear continua, 
and is consistent with a nuclear E(B-V)\,$<$\,1.1. 
But in the K-band, the continuum at the 
nucleus is much steeper (redder) than in the extranuclear spectra,
implying E(B-V)$\,\ge\,$3, in disagreement
with the low reddening of the J--band.

We also find that extrapolation of the J--band spectrum to the
K--band gives a continuum flux at $\lambda \approx 2.07\mu$m in
agreement with that observed.
At longer wavelengths, the spectrum is much redder,
suggesting we are seeing 
another component. According to models (e.g. Pier
\& Krolik 1993), this spectral region corresponds to the onset of
the emission from a dusty torus heated by the AGN. In fact, we
have successfully fitted a simple blackbody curve to the nuclear
K--band spectrum, but the temperature depends on the 
contribution of other components, such as 
the underlying galaxy spectrum. For example, subtracting a constant
contribution at the K--band of 85\% at $\lambda$2.15$\mu$m, for 
these components, we obtain a blackbody temperature of T = 730~K.
We thus conclude that the red nuclear continuum is most probably a result of
hot dust emission. This red continuum is only present
right at the nucleus, and is not resolved (size $\lesssim$ 150\,pc),
consistent with the torus hypothesis.

\subsection{Circinus}

\subsubsection{Emission-line Profiles}

Figure 8 shows the J and K spectra of the Circinus galaxy
obtained with the HR grating along p.a.\,=\,24\deg (close to the galaxy
major axis), while Fig. 9 shows the spectra along
p.a.\,=\,$-$66\deg (close to the radio axis). 
Emission in the lines [Fe\,II]$\lambda 1.257\mu$m, Pa$\,\beta$,
H$_2\,v$=1--0 S(1) and Br$\,\gamma$ extends beyond the central  
15$^{\prime\prime}$ covered by the slit, while
the coronal line of [S~IX] at $\lambda$1.252$\mu$m is spatially
unresolved ($\le$ 1\sec; 19 pc), 
as previously found by Oliva et al. (1994), and consistent
with modelling by Binette et al. (1997).

\begin{figure}
\vspace{10.0cm}
\includegraphics{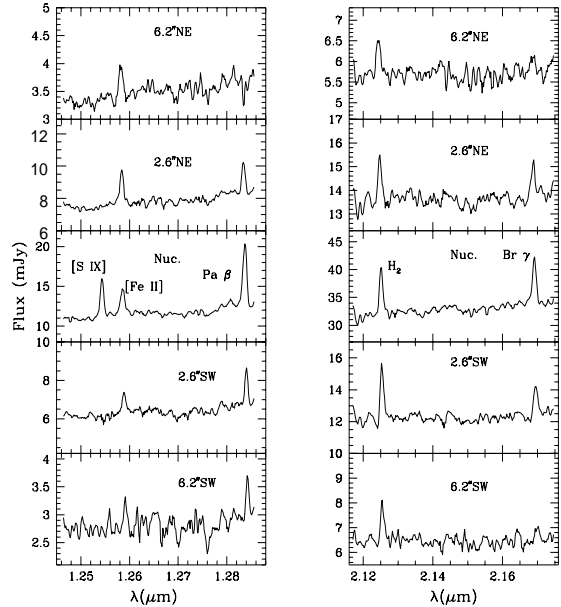}
\caption{J and K--band spectra of Circinus at $R\approx 2000$
along the galaxy plane at
 p.a.\,=\,24$^\circ$ after binning 2 pixels (0.73\sec) together. 
The distance from the nucleus is indicated.}
\label{fig8}
\end{figure}

\begin{figure}
\vspace{10.0cm}
\includegraphics{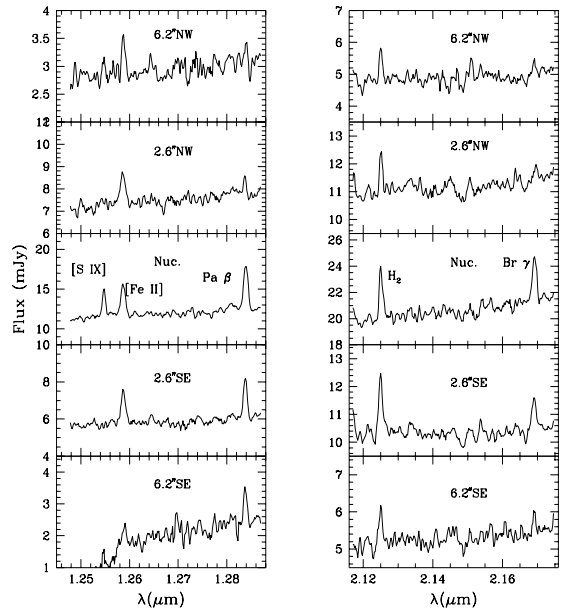}
\caption{J and K--band spectra of Circinus at $R\approx 2000$
along the radio axis at p.a.\,=\,$-$66$^\circ$ 
after binning 2 pixels (0.73\sec) together. 
The distance from the nucleus is indicated.}
\label{fig9}
\end{figure}

\begin{figure}
\vspace{6.0cm}
\includegraphics{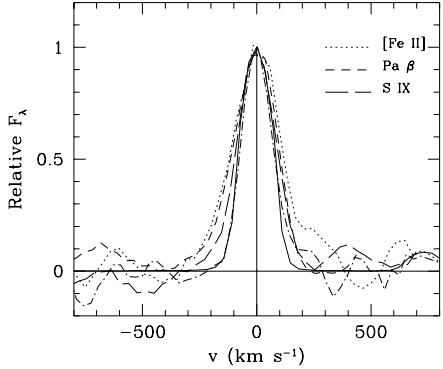}
\caption{Comparison of the nuclear emission-line profiles of Circinus with 
the instrumental profile (continuous line).}
\label{fig10}
\end{figure}

The emission lines are barely resolved spectrally, as illustrated in Fig. 10,
where the profiles of [S IX], [Fe II]$\lambda 1.257\mu$m and 
Pa$\,\beta$ in the nuclear J spectrum are shown together  
with the profile of a comparison lamp emission line. When corrected
by the instrumental profile, the FWHM of these emission lines are
97, 143 and 133~km~s$^{-1}$, respectively. The uncertainties are 
large (30--50 km s$^{-1}$) because the uncorrected FWHM values are close to the
instrumental value (150~km~s$^{-1}$). 
A common characteristic of all the nuclear emission-line profiles
is a slight blueward slanting asymmetry.

\subsubsection{Velocity field}

Figure 11 shows the heliocentric gas velocities of Circinus, obtained from
the peak wavelengths of [Fe II]$\lambda$1.257$\mu m$ and Pa$\,\beta$
along p.a.\,=\,24\deg and p.a.\,=\,$-$66\deg. Also shown is
the fit of a model with circular rotational
motions (Bertola et al. 1991) to the data along p.a.\,=\,24$^\circ$. 
As in the case of NGC\,2110,
the velocity field is dominated by circular rotation
in the plane of the galaxy, and  the kinematic centre
is displaced 1.0\sec\,SW of the peak of the IR continuum.
Our  data can be compared with IR and optical data from 
Maiolino et al. (1998), also shown in Fig. 11 as open squares: 
the inner 5 points correspond to stellar features in the K-band, 
and the outermost points
to optical emission in [NII]$\lambda$6584. Our results agree with
Maiolino et al.'s, filling nicely the gaps in their data.

\begin{figure}
\vspace{10.0cm}
\includegraphics{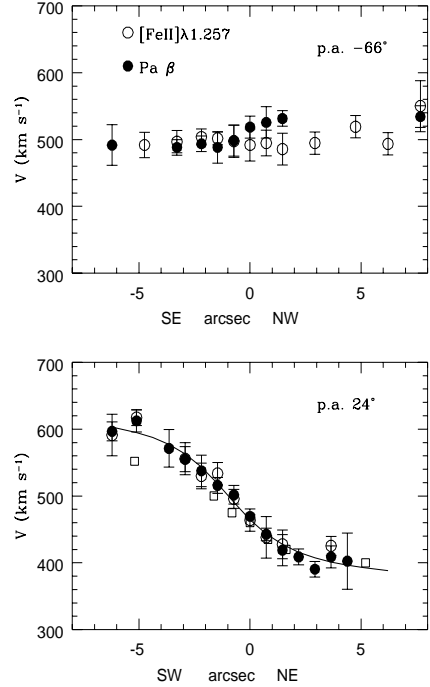}
\caption{Rotation curve of Circinus, obtained from the peak wavelength
of the emission-lines. The continuous line represents the fit of a 
model with circular rotational motions 
to the data along the galaxy plane at p.a.\,=\,24\deg. 
Squares represent data from
Maiolino et al. (5 inner points from K--band stellar absorption features
and 2 outermost points from the optical [NII]$\lambda$6584 emission line.)}
\label{fig11}
\end{figure}

\subsubsection{Emission line fluxes and ratios}

Figure 12 shows the line fluxes as a function of distance from the nucleus,
along p.a.\,=\,24\deg (major axis) and p.a.\,=\,$-$66$^\circ$ (radio axis). 
The [Fe II]$\lambda$1.257$\mu$m emission is, on average,
stronger along the radio axis than along the galaxy axis. 
The H$_2$ and Br$\,\gamma$
fluxes at the nucleus in the spectrum obtained along the radio axis
are about 50\% of the corresponding values in the spectrum obtained
along the galaxy major axis, which suggests that the slit
was not well centred in the K spectrum along p.a.\,=\,$-$66$^\circ$.

We have compared our flux distributions with the emission-line images of
Davies et al. (1998, hereafter D98) in  H$_2$ and [Fe\,II]$\lambda$1.64$\mu$m.
The latter line  comes from the same upper level as 
[Fe II]$\lambda$1.257$\mu$m (having intrinsically 75\% of its flux), and
we can thus compare the two brightness distributions. 

The H$_2$ brightness distribution of D98 shows
a central peak, corresponding to a marginally resolved central
source, with a steep radial profile,
superimposed on extended emission with a shallower
radial profile. From the lower left panel of Fig. 12, it can be seen that 
our H$_2$ fluxes present a similar behavior, showing a steeper
flux distribution within 2\sec from the peak (which seems to be
located $\approx$ 1\sec to SW and to SE of the peak in the continuum),
than in the outer parts.

The [Fe II] flux distribution of D98 is less centrally peaked,
showing an extension to the N--NE. 
Our [Fe II]$\lambda$1.257$\mu$m flux distribution 
consistently shows a ``flat-top'' along p.a. 24\deg. 
However, along the radio axis, our data is similarly
extended as along p.a. 24\deg, while D98 conclude 
that the [Fe II] emission is less extended towards the
cone.

The integrated H$_2\,v$=1--0,S(1) luminosity in the inner
$\approx 13$\sec (Fig. 12) is $4.3\times 10^{38}$~ergs~s$^{-1}$, 
which translates into a hot  H$_2$ mass of 140~M$_\odot$, 
using the same assumptions as for NGC 2110. Allowing for an
average reddening of E(B-V) $\approx $2.5 (see below), the above luminosity and
mass increase by a factor of $\approx $2.2.

\begin{figure}
\vspace{10.0cm}
\includegraphics{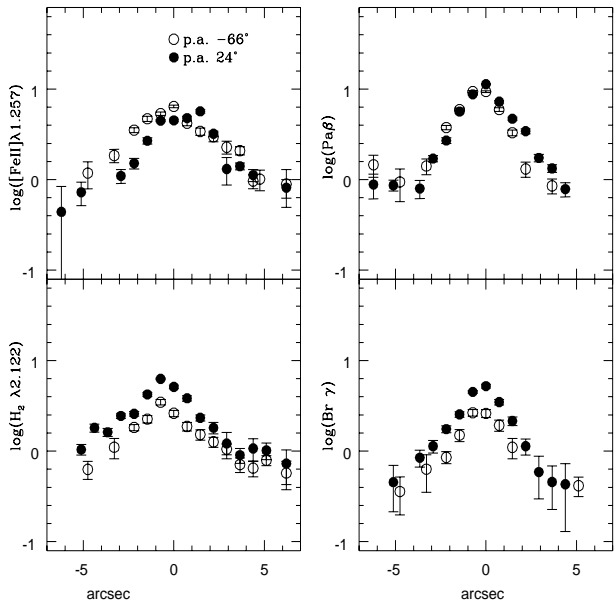}
\caption{Logarithm of the emission-line fluxes of Circinus (in units of
10$^{-15}$~ergs~cm$^{-2}$~s$^{-1}$~(arcsec)$^{-2}$) 
as a function of distance from the nucleus, along the 
galaxy plane at p.a.\,=\,24\deg 
(filled circles) and along the radio axis at
p.a.\,=\,$-$66\deg (open circles). For
p.a = 24$^\circ$, SW is negative and NE positive. For p.a.\,=\,$-$66$^\circ$, 
SE is negative and NW positive.}
\label{fig12}
\end{figure}

Figure 13 shows the line ratios
[Fe\,II]$\lambda$1.257/Pa$\,\beta$ and H$_2$/Br$\,\gamma$
along p.a.\,=\,24\deg and  p.a.\,=\,$-$66\deg.
[Fe\,II]/Pa$\,\beta$ has the value 0.4 at the nucleus and increases outwards,
most notably along the radio axis, reaching values larger than 2.
H$_2$/Br$\,\gamma$ presents a similar behaviour: it is $\approx 1$ at the nucleus and increases outwards to values larger than 2. 
Both ratios at the nucleus have values typical of starbursts (see discussion
above for NGC\,2110), suggesting that the starburst dominates the
gaseous excitation there. The larger ratios
away from the nucleus suggest that
X--ray radiation from the active nucleus and/or shocks 
dominate the excitation. In particular, the high
[Fe\,II]$\lambda$1.257/Pa$\,\beta$ ratios may trace the
high excitation gas, which extends along the galaxy minor axis
as shown by the ``ionization map'' (optical line ratio map 
[OIII]/(H$\alpha$+[NII]) of Marconi et al. (1994).

Our results regarding the nature of the H$_2$ emission
seem to contradict the interpretation
put forward by D98. They argue that the nuclear  H$_2$ component has the
flux expected by excitation produced by the observed
nuclear X-ray flux, and thus may originate from gas in the molecular
torus excited by the nuclear AGN, while the more 
extended emission could be due to on-going star formation.
Our results suggest the opposite is true (see Fig. 13, right panel).

\begin{figure}
\vspace{6.0cm}
\includegraphics{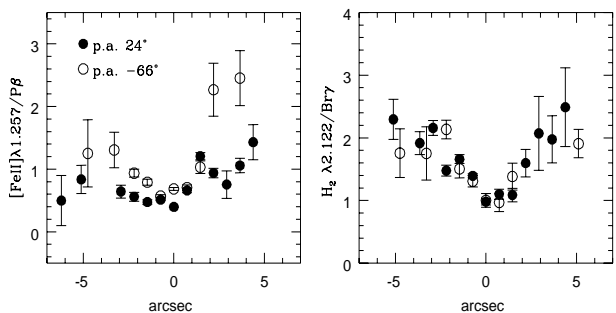}
\caption{Emission-line ratios of Circinus as a 
function of distance to the nucleus along the galaxy plane at
p.a.\,=\,24\deg (filled circles) 
and along the radio axis at
p.a.\,=\,$-$66\deg (open circles). Along p.a.\,=\,24$^\circ$, 
SW is negative and NE positive. Along p.a.\,=\,$-$66$^\circ$, 
SE is negative and NW positive.}
\label{fig13}
\end{figure}

\begin{figure}
\vspace{5.0cm}
\includegraphics{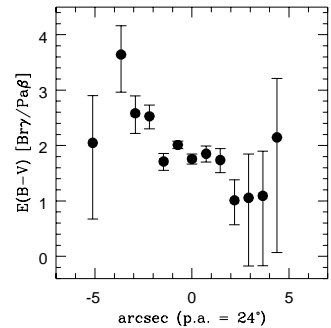}
\caption{Reddening E(B--V) (in mag)
as a function of distance from the nucleus
of Circinus along the galaxy plane at
p.a.\,=\,24$^\circ$. SW is negative and NE positive.}
\label{fig14}
\end{figure}

\subsubsection{Reddening and Dust Emission}

We have used the line fluxes in Pa$\,\beta$ and Br$\,\gamma$ to obtain
the gaseous reddening along p.a.\,=\,24$^\circ$ (we did not calculate
the reddening along p.a.\,=\,$-$66$^\circ$ due to the apparent misplacement
of the slit in the K-band spectrum at this p.a.). After 
correction for the foreground Milky Way reddening
E(B--V)$_G = 0.5$ mag (Freeman
et al. 1977), the internal reddening lies in the range 
1~$\le$~E(B--V)~$\le$~3.5 mag, and increases from NE to SW, as shown
in Fig. 14. 

The continuum colour (corrected for E(B--V)$_G$)
is reddest at the nucleus, where J--K$ = 1.9$, decreasing 
to J--K$ \approx 1.3$ to the NE, and to
J--K$ \approx$1.5--1.6 to the SW.

On the assumption
that the reddening derived from the emission lines can be applied to
the continuum, we obtain intrinsic colours J--K in the range 0.4--0.8 for
the extranuclear spectra and J--K$\,\approx 1$ for the nuclear spectrum.
The former J--K colours are typical of starbursts of ages 10$^6$--10$^8$ yrs
(Leitherer \& Heckman 1995). This result is in agreement with the 
modelling of Maiolino et al. (1998), who concluded that, on the basis of a 
Br $\gamma$ emission map,  within 100 pc from the
nucleus the stellar population has an age ranging between 4 $\times 10^7$ 
and 1.5 $\times 10^8$ yrs.

\section{Summary and Concluding Remarks}

Long-slit spectra in the near IR J and K--bands of the Seyfert 2 galaxies 
NGC\,2110 and Circinus  have revealed extended emission in 
[Fe\,II]$\lambda$1.257$\mu$m, Pa$\,\beta$ and
H$_2\,v$=1--0 S(1) up to at least $\approx$ 900 pc (6\sec)
from the nucleus in NGC\,2110 and beyond the end of the slit
-- 130 pc (7\sec) from the nucleus -- in the case of Circinus.

The profiles of the emission lines [Fe\,II]$\lambda 1.257\mu$m and
Pa$\,\beta$ from the nucleus of NGC\,2110 are broad 
(FWHM $\approx$ 500 km s$^{-1}$) 
and quite  similar, while the profile of H$_2\,v$=1--0 S(1)
is narrower, with FWHM$\le$300 km s$^{-1}$,
and thus originates in a kinematically less disturbed gas.
We do not confirm previous reports of a very
broad Pa$\,\beta$ component. 
In Circinus, the profiles are barely resolved spectrally, with
FWHM $\le$ 150 km s$^{-1}$.

The H$_2$ line luminosities and masses of hot molecular gas 
are similar to those of other
Seyfert galaxies (e.g. Veilleux et al. 1997).
The [Fe\,II]$\lambda 1.257\mu$m emission may trace the high excitation gas.
In the case of NGC\,2110, the high [Fe\,II]$\lambda 1.257\mu$m/Pa$\,\beta$ flux 
ratio combined with the broadening of the nuclear
profiles suggests that shocks (perhaps driven by the radio
jet) are an important source of excitation of the [Fe\,II] emission.
The high H$_2\,v$=1--0 S(1)/Br$\,\gamma$ nuclear ratio,
combined with the smaller width of the H$_2$ emission--line,
suggests that the excitation of the H$_2$ line is dominated by
X--ray emission from the active nucleus. In the case of
Circinus, the much lower ratios at the nucleus are similar
to those observed for starbursts, suggesting that the nuclear
starburst is the main source of excitation there. 
However, a few arcseconds from the nucleus, these ratios increase
to values similar to those found in Seyfert nuclei, 
showing that the radiation of the active nucleus
may be the main source of excitation outside the nucleus.   

We were able to obtain rotation curves in the IR lines
for both galaxies, which indicate that the gaseous kinematics
are dominated by circular motions in the disks of the galaxies.
In both cases we found a displacement between the peak of the
IR continuum and the kinematic center of the galaxy along the
p.a. closest to the major axis. 
In the case of NGC\,2110, this effect had been previously found  using
optical observations, but the IR rotation curve is more symmetric 
relative to the nucleus than the rotation curve obtained from optical
lines. This result suggests that the offset of the continuum
nucleus with respect to the kinematic centre is, at least in part, 
an effect of obscuration.

For NGC\,2110, J--K~$\approx$~0.9 everywhere, except right at
the nucleus, where J--K~$\approx$~1.9. The continuum spectrum within the
K band is very steep and cannot be explained by reddening alone. 
The observations clearly
show that we are seeing another component in emission in the K--band,
which is consistent with the expected emission of dust heated by an AGN. 
This emission is unresolved in our data (0.73\sec$ \times $1.1\sec), 
and may originate in the walls of a circumnuclear dusty torus. 

For Circinus, the Br~$\gamma$/Pa~$\beta$ ratios indicate gaseous
reddening in the range 1~$\le$~E(B-V)~$\le$~3. The J--K values
are consistent with those of an ageing starburst with the same reddening
as the emitting gas, except at the unresolved
nucleus, which has a redder colour.

\section*{acknowledgements}
We thank the support of the staff of CTIO by assistance with the
near--IR observations and reductions, in particular Richard Elston.
We also thank the referee, Chris Done, for several 
suggestions which helped to improve the paper.
This work was partially supported by the Brazilian institutions
CNPq, FAPERGS and FINEP. This research has made use of the NASA/IPAC
Extragalactic Database (NED), which is operated by the Jet Propulsion
Laboratory, under contract with NASA.

\end{document}